\documentclass[aip,jcp,amsmath,amssymb,reprint,groupedaddress]{revtex4-1}
\usepackage{graphicx}
\usepackage{amsmath}
\usepackage{mhchem}
\usepackage{braket}
\usepackage{multirow}
\usepackage{subcaption}
\usepackage{xcolor}
\usepackage{comment}
\usepackage{bm}
\usepackage{soul} 

\newcommand\noperiod[1]{}

\DeclareMathOperator{\Tr}{Tr}

\DeclareMathOperator{\Imag}{Im}
\newcommand{\iu}{\ensuremath{\mathrm{i}}}
\newcommand{\eu}{\mathrm{e}^}

\newcommand{\rmd}{\mathrm{d}}
\newcommand{\half}{{\ensuremath{\frac{1}{2}}}}
\newcommand{\quarter}{{\ensuremath{\frac{1}{4}}}}
\newcommand{\thalf}{{\ensuremath{\tfrac{1}{2}}}} 
\newcommand{\tquarter}{{\ensuremath{\tfrac{1}{4}}}}
\newcommand{\eqn}[1]{Eq.~\eqref{#1}}

\newcommand{\red}[1]{{\color{red}{#1}}{}} 

\newcommand{\pder}[3][]{\frac{\partial^{#1}{#2}}{\partial{#3}^{#1}}}

\newcommand{\Ref}[1]{Ref.~\onlinecite{#1}}
\usepackage{dcolumn} 
\newcolumntype{d}[1]{D{.}{.}{#1}} 
\newcommand{\ctitle}[1]{\multicolumn{1}{c}{#1}}

\begin{document}
	
	\title{Instanton theory of tunnelling in molecules with asymmetric isotopic substitutions}
	\author{Elena Jahr}
	\altaffiliation{These authors contributed equally}
	\author{Gabriel Laude}
	\altaffiliation{These authors contributed equally}
	\author{Jeremy O. Richardson}
	\email{jeremy.richardson@phys.chem.ethz.ch}
	\affiliation{Laboratory of Physical Chemistry, ETH Z\"urich, 8093 Z\"urich, Switzerland}
	
	\begin{abstract}
		We consider quantum tunnelling in asymmetric double-well systems for which the local minima in the two wells have the same energy, but the frequencies differ slightly.  {In a molecular context, this situation can arise if the symmetry is broken by isotopic substitutions}.
		We derive a generalization of instanton theory for these asymmetric systems,
		leading to an semiclassical expression for the tunnelling matrix element and hence the energy-level splitting.
		We benchmark the method using a set of one- and two-dimensional models, for which the results compare favourably with numerically exact quantum calculations.
		Using the ring-polymer instanton approach, we apply the method to compute the level splittings in various {isotopomers} of malonaldehyde in full dimensionality and analyse the relative contributions from the zero-point energy difference and tunnelling effects.
	\end{abstract}
	
	\maketitle
	
	\section{Introduction}
	
	The phenomenon of quantum tunnelling in a symmetric double well is a well-known concept. \cite{Landau+Lifshitz,BellBook}
	The earliest molecular example studied was the splitting of the ground-state energy levels of the ammonia molecule due to inversion tunnelling. \cite{Hund1927tunnel,*Dennison1932WKB,*Dennison1932NH3}
	However, the prototypical system of intramolecular hydrogen-atom tunnelling is malonaldehyde, \cite{Baughcum1981malonaldehyde,Baughcum1984malonaldehyde}
	whose ground-state tunnelling splitting has been measured to high accuracy
	\cite{Firth1991malonaldehyde,*Baba1999malonaldehyde}
	and studied with a wide variety of theoretical methods.
	\cite{Bicerano1983malonaldehyde,*Carrington1986malonaldehyde,*Ruf1988malonaldehyde,
		Coutinho2004malonaldehyde,Schroeder2011malonaldehyde,Hammer2011malonaldehyde,manthe2012,SchroederMCTDH2014,
		Tew2014PES,
		Wang2008malonaldehydePES,Wang2008malonaldehydeSplit,*Wang2013Qim,Burd2020tunnel,
		Yagi2001malonaldehyde,Tautermann2002malonaldehyde,
		Milnikov2003,*Milnikov2004,
		tunnel,MUSTreview,Cvitas2016instanton,Cvitas2018instanton,Erakovic2020instanton}
	
	
	Instanton theory, \cite{Miller1975semiclassical,Uses_of_Instantons,ABCofInstantons,Benderskii}
	which is based on a semiclassical approximation to the path-integral formulation of quantum mechanics, \cite{Feynman,Kleinert}
	provides a particularly powerful approach for computing tunnelling splittings,
	especially
	when reformulated into the 
	ring-polymer instanton method \cite{tunnel,Perspective,InstReview}
	or related approaches. 
	\cite{Milnikov2001,*Milnikov2008review,Cvitas2016instanton,Cvitas2018instanton,Erakovic2020instanton}
	These methods have been applied to calculate the ground-state tunnelling splitting in numerous molecular systems in full dimensionality.
	\cite{tunnel,formic,water,octamer,hexamerprism,MUSTreview,i-wat2,pentamer,WaterChapter,Kawatsu2015NH3,
		Milnikov2003,*Milnikov2004,Milnikov2006vinyl,Cvitas2016instanton,Cvitas2018instanton,Erakovic2020instanton}
	
	In contrast, there has been much less theoretical work on tunnelling between asymmetric wells.%
	\footnote{Note that tunnelling pathways in water clusters commonly have asymmetric profiles although they still link equivalent wells and can thus be studied with the standard instanton theory.\cite{water,Erakovic2020instanton}}
	Some exceptions include studies of tunnelling between similar (but not equivalent) minima in the water hexamer cage.
	\cite{Liu1996hexamer,*Liu1997cage,*Gregory1997hexamer,Walesdimerhexamer}
	However, in this paper, we shall study a subtly different problem,
	one in which both well minima have exactly the same potential energy, but different frequencies.
	This can occur (within the Born--Oppenheimer approximation) 
	in molecules or clusters with certain asymmetric isotopic substitutions.
	Previous studies of 
	tunnelling switching in meta-D-phenol \cite{Albert2013phenol,*Albert2016metaDphenol}
	and the energy-level structure in the dimer \ce{HF-DF} \cite{Quack1990HF2,*Farrell1996HFDF,*He2007HF2}
	provide examples which exhibit this behaviour.
	In both these cases, however, the 
	effect of tunnelling between the ground vibrational states of each well
	was predicted to be much smaller than the zero-point energy (ZPE) difference between the two wells.
	
	In this work, we study cases where the asymmetry is weak enough such that
	tunnelling may still have a significant effect on the ground-state energy levels of the molecule. 
	We develop a theory which can be applied using ring-polymer instanton methodology \cite{tunnel,Perspective,InstReview} to compute the effect of tunnelling on the level splitting in such molecular systems.
	This theory is based on minimum-action tunnelling pathways, called instantons, which connect the bottom of the wells.
	This is in contrast to 
	alternative approaches which go by various names
	such as WKB theory, \cite{Landau+Lifshitz,Garg2000,Song2015WKB,Halataei2017WKB}
	semiclassical path integral, \cite{Holstein1988tunnelling}
	Bohr--Sommerfield quantization \cite{Child,Chebotarev1998WKB} or periodic-orbit theory,\cite{Miller1979periodic}
	for which the tunnelling pathways travel between two turning points defined by the energy of a localized vibrational state.
	These approaches may give good results for model systems,
	but unlike instanton theory, they cannot be easily extended to multidimensional problems (without \emph{a priori} choices of path such as that of \Ref{Makri1989tunnel}).
	
	
	We will apply our new instanton approach to benchmark one- and two-dimensional model systems in various parameter regimes and thereby test its accuracy.
	Then, we apply it in full dimensionality to compute the level splitting in a set of 
	isotopomers of malonaldehyde, including both symmetric and asymmetric substitutions
	and compare with experimental results where available. \cite{Baughcum1981malonaldehyde,Baughcum1984malonaldehyde}

	

	\section{Theory}
	\label{sec:theory}
	
	\begin{figure}
		\includegraphics[width=\linewidth]{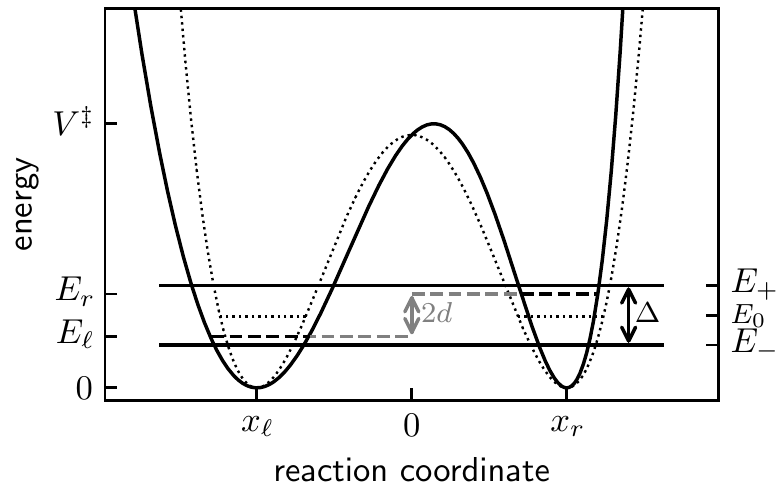}
		\caption{Schematic representation of the potential, $V(x)$,
			for an asymmetric system (solid curve) with non-tunnelling zero-point energy levels shown with horizontal dashed lines.
			For comparison the potential of a symmetric system is shown with the grey dotted curve with its non-tunnelling zero-point energy levels $E_0$ (horizontal grey dotted lines).
			The solid horizontal lines indicate the lowest two levels of the asymmetric system including tunnelling.
		}
		\label{fig:potential}
	\end{figure}
	
	
	We study tunnelling in a system defined by the Hamiltonian
	$\hat{H} = \hat{p}^2/2m + V(\hat{x})$.
	This expression is valid for both one-dimensional and multidimensional systems,
	as in the case of the latter we treat position, $x$, and momentum, $p$, as vectors
	which have been mass-weighted such that all degrees of freedom have the same effective mass, $m$.
	We will consider potentials, $V(x)$, which
	exhibit a special type of asymmetric double-well structure, as
	illustrated in Fig.~\ref{fig:potential}.
	In particular, there are two minima, at $x_\ell$ and $x_r$, which have exactly the same potential energy and which we choose to set to zero, i.e.\ $V(x_\ell)=V(x_r)=0$.
	However, the frequencies, 
	$\omega_{\ell/r}=\sqrt{\nabla^2V(x_{\ell/r})/m}$,
	in the two wells may differ 
	leading to different ZPE values,
	$E_\ell\simeq\thalf\hbar\omega_\ell$ and $E_r\simeq\thalf\hbar\omega_r$ (within the harmonic approximation).
	\footnote{In the multidimensional case, the normal-mode frequencies are defined as the square root of the eigenvalues of the mass-weighted Hessian matrices and the ZPE is defined by a sum over modes.} 

	For such a system, we can define an effective two-level Hamiltonian 
	using a basis of states localized in the left and right wells,
	\begin{align}
	\label{eq:asym-H}
	H_\text{eff} = \begin{pmatrix}
	E_0 - d & -\hbar\Omega \\
	-\hbar\Omega & E_0 + d
	\end{pmatrix},
	\end{align}
	where $d=\frac{1}{2}(E_r - E_\ell)$ and $E_0 = \frac{1}{2}(E_\ell + E_r)$.
	Here, $-\hbar\Omega$ is the coupling due to tunnelling between the left and right wells,
	which will be evaluated by the instanton theory developed in this paper.
	By introducing this effective Hamiltonian, we have implicitly assumed that the asymmetry is weak enough such that 
	the ground vibrational state of the left well couples to the ground vibrational state of right well and we have neglected couplings to other levels.%
	\footnote{We have also neglected the overlap between wavefunctions in the left and right wells.  This is valid so long as the barrier is high or wide, which is anyway a requirement for the semiclassical approach.
		The effective Hamiltonian is defined by the matrix elements of the full Hamiltonian in this basis
		and its eigenvalues are solutions of the secular equations obtained from the variational method given a trial wavefunction defined by a linear combination of the basis states. 
	}
	This ensures that the tunnelling will remain coherent, which is the regime of interest as
	an instanton theory for incoherent tunnelling (i.e.\ rate constants) is of course already well known.
	\cite{Miller1975semiclassical,Uses_of_Instantons,Benderskii,Andersson2009Hmethane,RPInst,Althorpe2011ImF,AdiabaticGreens,Perspective,InstReview}
	
	Simple quantum-mechanical principles based on the effective Hamiltonian of \eqn{eq:asym-H}
	allow for a reduced description of the dynamics and spectrum of the full system.
	The solutions are well known as this
	is an equivalent problem for all two-level quantum systems, \cite{FeynmanLecturesIII}
	e.g.\ molecular orbitals defined in terms of a linear combination of atomic orbitals in a heteronuclear diatomic or a spin-\thalf\ in a magnetic field.
	The effective Hamiltonian has eigenvalues $E_\pm = E_0 \pm \sqrt{d^2 + (\hbar\Omega)^2}$,
	and thus
	for an energy-resolved spectrum
	the observable of interest is the level splitting,
	$\Delta = 2\sqrt{d^2 + (\hbar \Omega)^2}$,
	which has contributions both from the tunnelling effect as well as the ZPE difference.
	The ground- and excited-state eigenvectors of the Hamiltonian matrix can be written as $\psi_-=(\cos\frac{\phi}{2},\sin\frac{\phi}{2})$ and $\psi_+=(-\sin\frac{\phi}{2},\cos\frac{\phi}{2})$,
	where $\phi = \tan^{-1}(\hbar\Omega/d)$ is the mixing angle ($0\le\phi\le180^\circ$).
	This angle is
	a measure of localization of the stationary states and
	is a function only of the ratio $d/\hbar\Omega$.
	That is, for mixing angle $\phi=90^\circ$, the eigenstates are maximally delocalized (with a population of $\thalf$ in each well), but if the angle approaches 0 or $180^\circ$, the ground- and excited-state wavefunctions are each localized to one well.
	One can also obtain the time-dependence of the populations from the effective Hamiltonian.
	For example if the system starts in the left well, the time-dependent population in the right well is 
	$\sin^2\phi \, \sin^2(\Delta t/2\hbar)$
	and thus the mixing angle, $\phi$, also determines the amplitude of population transfer.
	This analysis shows that one requires only the values of $d$ and $\Omega$ to make predictions for all these observables of interest.
	It is easy to obtain a good estimate of $d$ via the ZPEs in each well evaluated using the harmonic approximation as described above.
	However, the tunnelling contribution, $\Omega$, cannot be obtained from a harmonic approximation
	\footnote{In fact, even perturbation theory around the well minimum is not sufficient to describe these tunnelling effects.\cite{Polyakov1977instanton}  The instanton approach instead captures the exponentially small tunnelling effect using an asymptotic series in $\hbar$. \cite{Kleinert}\protect\noperiod.}
	and requires a more involved treatment, to which we dedicate the rest of this section.
	
	As has been done for the symmetric case, \cite{Benderskii,tunnel,InstReview}
	we can start our derivation by considering the partition function of 
	the effective Hamiltonian [\eqn{eq:asym-H}], which is
	$Z = 2\,\eu{-\beta E_0} \cosh(\beta \sqrt{d^2 + (\hbar \Omega)^2})$,
	where $\beta$ is the reciprocal temperature.
	%
	%
	However, we must generalize the standard approach in order to have an appropriate reference partition function for the non-tunnelling case.
	Here, we define $Z_\text{g} = 2\sqrt{Z_\ell Z_r}$ in terms of the geometric average of  
	$Z_\ell = \eu{-\beta(E_0-d)}$ and $Z_r = \eu{-\beta(E_0+d)}$, the partition functions for 
	the left/right well respectively (i.e.\ in the absence of tunnelling, where $\Omega=0$), such that $Z_\text{g} = 2\,\eu{-\beta E_0}$.
	The ratio of the total partition function to this reference is
	\begin{align}
	\label{eq:Z_ratio}
	\frac{Z}{Z_\text{g}} &= \cosh(\beta \sqrt{d^2 + (\hbar \Omega)^2}),
	\end{align}
	which for later convenience can be equivalently written as a Taylor series in $\Omega$:
	\begin{align}
	\label{Zratio}
	\frac{Z}{Z_\text{g}} &= \cosh(\beta d) + \frac{\beta \hbar^2\Omega^2 \sinh(\beta d)}{2d} \nonumber \\
	& \quad\qquad + \frac{\beta \hbar^4\Omega^4 \left[\beta d\cosh(\beta d) - \sinh(\beta d)\right]}{8d^3} + \cdots.
	\end{align}
	
	
	The effective Hamiltonian only defines the two lowest states of the system.
	Thus, writing the partition function in terms of this two-level Hamiltonian
	employs an approximation which is only valid if the temperature is chosen such that
	the thermal energy ($\beta^{-1}$) is much smaller than the spacing between the vibrational states of a single well ($\hbar\omega_{\ell/r}$).
	In Sec.~\ref{subsec:instanton}, we will therefore take the low-temperature, $\beta\rightarrow\infty$, limit of the full partition function in order that it corresponds with the expressions based on $H_\text{eff}$.

	\subsection{Instanton Theory}
	\label{subsec:instanton}
	
	
	To derive an method for calculating $\Omega$ and hence $\Delta$ within instanton theory, we require expressions for $Z$ and $Z_\text{g}$ based on the full Hamiltonian, $\hat{H}$, which we can compare with the previous analysis based on $H_\text{eff}$.
	The partition function $Z=\Tr[\eu{-\beta\hat{H}}]$ can be defined within the path-integral formalism as a sum over all closed paths in the space of $x$. \cite{Feynman}
	Some of these paths are located completely in one of the wells, whereas others pass back and forth 
	between the wells $n$ times.
	Each pass is called a ``kink'' and in order for the path to be closed in this simple double-well case, $n$ must be an even number.
	We define $Z_n$ as the contribution to the partition function from paths with $n$ kinks such that \cite{Benderskii,tunnel,InstReview}
	\begin{align} \label{Zkinks}
	\frac{Z}{Z_\text{g}} = \frac{Z_0}{Z_\text{g}} + \frac{Z_2}{Z_\text{g}} + \frac{Z_4}{Z_\text{g}} + \cdots    + \frac{Z_n}{Z_\text{g}} + \cdots.
	\end{align}
	We will evaluate semiclassical approximations to $Z_n$ for each value of $n$,
	which are rigorously defined by taking the leading term of the asymptotic expansion in $\hbar$.
	{For this, we will use the notation $A\simeq B$ to indicate that a function $B$ is an asymptotic approximation \cite{BenderBook}
		to $A$ and that therefore $B/A\rightarrow1$ 
		for low temperature or long imaginary time
		in the limit $\hbar\rightarrow0$.}
	
	Consider first the $Z_0$ term, which includes contributions from all non-tunnelling paths
	completely located either in the left or right wells.
	We define $K_{0}(x_a,x_b,\tau)$ as the zero-kink contribution to the semiclassical approximation of $\braket{x_a|\eu{-\tau\hat{H}/\hbar}|x_b}$. \cite{InstReview}
	Then we define $Z_\ell$ by
	taking the integral only over the domain of the left well, $\Gamma_\ell$,%
	\footnote{Note that as long as this domain is continuous and includes $x_\ell$ but not $x_r$, the result of the semiclassical approximation will be independent of the exact definition of $\Gamma_\ell$.  A simple choice would be 
		the interval $(-\infty,0)$.}
	\begin{align}
	Z_\ell &= \int_{\Gamma_\ell} \rmd x_a \, K_0(x_a,x_a,\beta\hbar)
	\simeq \Xi_\ell K_\ell(\beta\hbar) ,
	\end{align}
	and equivalently for $Z_r$,
	where $K_\ell(\tau) = K_0(x_\ell,x_\ell,\tau)$
	and 
	$K_r(\tau) = K_0(x_r,x_r,\tau)$ are the kernels for a path starting and ending at the bottom of one of the wells without crossing the barrier,
	whereas $\Xi_\ell$ and $\Xi_r$ are terms resulting from the steepest-descent integration over the end points.
	Explicit formulas are given in Appendix~\ref{app:HO} which can be used to show that according to the semiclassical approximation,
	these partition functions are
	\begin{align}
	Z_\ell &\simeq \eu{-\half\beta\hbar\omega_\ell} &     
	Z_r &\simeq \eu{-\half\beta\hbar\omega_r} ,
	\end{align}
	as could have been anticipated from a simple harmonic analysis in the low-temperature limit.
	Using $Z_0 = Z_\ell + Z_r$ and $Z_\text{g} = 2\sqrt{Z_\ell Z_r}$,
	the first term in \eqn{Zkinks} gives
	\begin{align}
	\frac{Z_0}{Z_\text{g}} &\simeq \frac{\eu{-\half\beta\hbar\omega_\ell} + \eu{-\half\beta\hbar\omega_r}}{2\,\eu{-\quarter\beta\hbar(\omega_\ell+\omega_r)}}
	= \cosh\left(\tquarter\beta\hbar(\omega_r - \omega_\ell)\right).
	\end{align}
	Comparing this with the first term in \eqn{Zratio}, we can identify $d=\tquarter\hbar(\omega_r - \omega_\ell)$ as expected.
	
	In this way our path-integral analysis of the non-tunnelling pathways has presented us with a formula for calculating $d$.
	There are of course simpler methods which give the same result, but the advantage of using the path-integral formulation will come when considering the remaining ($n>0$) terms in the series.
	
	
	Generalizing the approach of Ref.~\citenum{InstReview},
	we can split the paths which contribute to $Z_n$ into $n$ kinks.
	To define the contribution from a single kink, we first consider the semiclassical limit of the propagator 
	$\braket{x_\ell|\eu{-\tau_\text{k}\hat{H}/\hbar}|x_r}$.
	The paths which dominate this matrix element are single kinks of length $\tau_\mathrm{k}$
	which start in the left well and end in the right.
	We will focus on a particular dominant path which is centred such that it passes through a point of maximum potential at imaginary time $\tau_\text{k}/2$.
	We call this dominant path an ``instanton''; it is a minimum-action pathway and is thus equivalent to a classical trajectory in imaginary time. \cite{Miller1971density,Uses_of_Instantons}
	Including also the non-zero fluctuations around this path leads to an instanton contribution which we call
	$K_1'(x_\ell,x_r,\tau_\text{k})$, 
	where the prime indicates the restriction that the path be centred as described above.
	This quantity is defined explicitly in Eq.~(90) of Ref.~\citenum{InstReview},
	and can be evaluated within the ring-polymer instanton formalism using \eqn{K1}.

\begin{figure}[t!]
		\includegraphics[width=\linewidth]{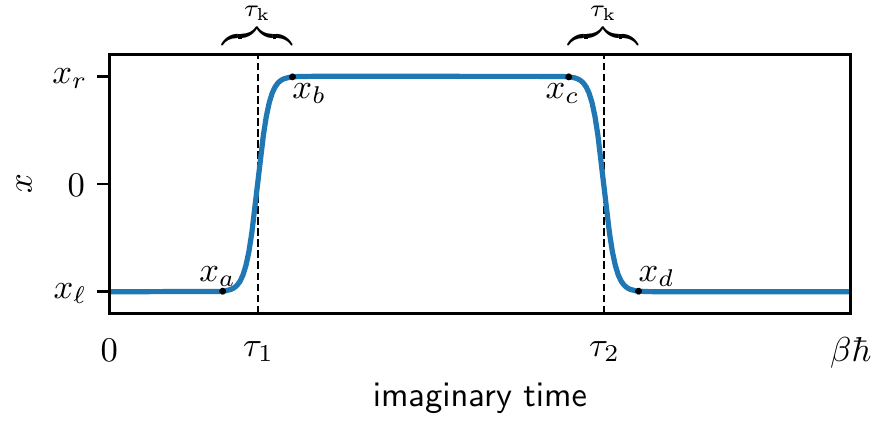}
		\caption{Schematic of a 2-kink path starting and finishing in the left well in total imaginary time $\beta\hbar$.
			The kinks are located within a relatively short width $\tau_\mathrm{k}$ and are centred at imaginary times $\tau_1$ and $\tau_2$.
		}
		\label{fig:kinks}
	\end{figure}

	\begin{widetext}
		We are now in a position to evaluate contributions to the two-kink term.
		Here, it is again necessary to extend the standard theory so as to keep track of the amount of imaginary time spent by the path trapped in each of the wells, which are no longer equivalent.
		The path is broken up into four pieces with 
		two kinks of length $\tau_\mathrm{k}$ centred around imaginary times $\tau_1$ and $\tau_2$ as depicted in Fig.~\ref{fig:kinks}.
		Therefore,
		\begin{subequations}
			\begin{align}
			Z_2'(\tau_1,\tau_2) &= \int_{\Gamma_\ell} \rmd x_a \int_{\Gamma_r} \rmd x_b \int_{\Gamma_r} \rmd x_c \int_{\Gamma_\ell} \rmd x_d
			\nonumber\\&\qquad\times 
			K_1'(x_a,x_b,\tau_\text{k}) K_0(x_b,x_c,\tau_2-\tau_1-\tau_\text{k})
			K_1'(x_c,x_d,\tau_\text{k}) K_0(x_d,x_a,\beta\hbar-\tau_2+\tau_1-\tau_\text{k})
			\nonumber\\&\qquad + [\text{the same with } \ell\Longleftrightarrow r]
			\\ &\simeq \Xi_\ell K_1'(\tau_\text{k}) \Xi_r K_r(\tau_2-\tau_1-\tau_\text{k}) \Xi_r K_1'(\tau_\text{k}) \Xi_\ell K_\ell(\beta\hbar-\tau_2+\tau_1-\tau_\text{k})
			\nonumber\\&\qquad+ 
			\Xi_r K_1'(\tau_\text{k}) \Xi_\ell K_\ell(\tau_2-\tau_1-\tau_\text{k}) \Xi_\ell K_1'(\tau_\text{k}) \Xi_r K_r(\beta\hbar-\tau_2+\tau_1-\tau_\text{k}) ,
			\end{align}
		\end{subequations}
		where we have defined $K_1'(\tau_\text{k}) = K_1'(x_\ell,x_r,\tau_\text{k}) = K_1'(x_r,x_\ell,\tau_\text{k})$, which are equal due to time-reversal symmetry.
		Finally, using the formulas in Appendix~\ref{app:HO}, we obtain
		\begin{align}
		Z_2'(\tau_1,\tau_2)
		&\simeq \theta(\tau_\text{k})^2 \left[ \eu{-\half\omega_r(\tau_2-\tau_1)} \eu{-\half\omega_\ell(\beta\hbar-\tau_2+\tau_1)}
		+ \eu{-\half\omega_\ell(\tau_2-\tau_1)} \eu{-\half\omega_r(\beta\hbar-\tau_2+\tau_1)} \right] ,
		\label{Z2prime}
		\end{align}
	\end{widetext}
	where we have defined
	\begin{align} \label{eq:theta}
	\theta(\tau_\text{k}) = \frac{K_1'(\tau_\text{k})}{\sqrt{K_\ell(\tau_\text{k})K_r(\tau_\text{k})}} .
	\end{align}
	The formulas from Appendix~\ref{app:HO} are only valid in the long-time limit and thus we must choose $\tau_\mathrm{k}$ large enough that this expression converges.
	However, the value of $\beta\hbar$ is assumed to be much larger again such that the kinks are widely spaced and interactions between the kinks can be neglected.

	In order to obtain the full expression for $Z_2$, one needs to integrate over the imaginary times at which the kinks occur, 
	which is simple because $\theta(\tau_\text{k})$ does not depend on these variables:%
	\footnote{Here we are using a similar argument to that used for the standard assumption of a non-interacting instanton gas, which is valid so long as the density of kinks does not become so high that they overlap.\cite{Uses_of_Instantons,Kleinert}
		This in turn is valid for systems with high barriers and low probability of tunnelling.
		Actually in our case, $Z_2'(\tau_1,\tau_2)$ does depend on the locations of the kinks even though it ignores their interaction and thus the partition function does not correspond to a gas of kinks, but instead to a one-dimensional spin lattice with domain walls located at the kinks. \cite{Polyakov1977instanton,ChandlerGreen}
		In this analogy, the necessary asymmetry would be provided by a weak magnetic field.}
	\begin{subequations}
		\begin{align}
		Z_2 &= \int_0^{\beta\hbar} \rmd \tau_1 \int_{\tau_1}^{\beta\hbar} \rmd \tau_2 \, Z_2'(\tau_1,\tau_2)
		\\ &\simeq 2\beta\hbar \theta(\tau_k)^2 \frac{\eu{-\half\beta\hbar\omega_\ell} - \eu{-\half\beta\hbar\omega_r}}{\omega_r-\omega_\ell} .
		\end{align}
	\end{subequations}
	Therefore the required ratio is
	\begin{align}
	\frac{Z_2}{Z_\text{g}} &\simeq 2\beta\hbar\theta(\tau_k)^2 \frac{\sinh\left(\quarter\beta\hbar(\omega_r - \omega_\ell)\right)}{\omega_r - \omega_\ell} ,
	\end{align}
	with $\tau_\mathrm{k}\rightarrow\infty$ implied.
	In order that this matches the second term of \eqn{Zratio} we must again identify $d=\tquarter\hbar(\omega_r - \omega_\ell)$ as before.
	However, we also gain an extra piece of information which implies that 
	\begin{align}
	\Omega \simeq \lim_{\tau_\text{k}\rightarrow\infty} \theta(\tau_\text{k}) ,
	\end{align}
	and thereby defines the instanton approximation for the tunnelling matrix element.

	With these definitions, 
	the agreement between the two series continues if one computes $Z_4/Z_\mathrm{g}$ etc., as of course it should due to the fact that the semiclassical approximation recovers the asymptotic ($\hbar\rightarrow0$) limit.
	This can also be understood by considering the 
	perturbation-theory expansion for $Z=\Tr[\eu{-\beta H_\text{eff}}]$, 
	which generates \eqn{Zratio} directly and
	has terms of a similar structure to those in \eqn{Z2prime} and its $n\ge2$ generalizations.
	\footnote{To do this, split the effective Hamiltonian into diagonal and off-diagonal parts: $H_\text{eff} = H_\text{eff}^{(0)} + H_\text{eff}^{(1)}$.
		The perturbation series can then be obtained using the recursive formula:
		\[ \eu{-\beta H_\text{eff}} = \eu{-\beta H_\text{eff}^{(0)}} + \tfrac{1}{\hbar} \int_0^{\beta\hbar} \! \eu{-\tau H_\text{eff}/\hbar} H_\text{eff}^{(1)} \eu{-(\beta\hbar-\tau) H_\text{eff}^{(0)}/\hbar} \, \rmd \tau . \]
		\protect\noperiod.
	}
	Alternatively, one can see that the correct expansion must be recovered because of the well-known isomorphism between a two-level system and an infinite periodic one-dimensional Ising model\cite{ChandlerGreen}
	as well as the less-well-known isomorphism between the latter and the non-interacting kinks. \cite{Polyakov1977instanton}

	Finally, we can check that our assumption of non-interacting kinks remains valid for the total partition function.
	In the standard instanton theory for symmetric systems, \cite{Uses_of_Instantons}
	one can predict that the average number of kinks which will appear is $\beta\hbar\Omega$, which is typically low enough such that the intervals between them ($\Omega^{-1}$) is much larger than the width of the kinks ($\approx2/\omega_0$, where $\omega_0$ is the frequency of the wells).\cite{Kleinert}
	It can be shown by similar arguments that the average number of kinks in an asymmetric system is less than or equal to that in the symmetric case such that the proof holds also for our case.%
	\footnote{In the asymmetric case, the average number of kinks is
		\[\braket{n} = \frac{1}{Z}\sum_n n Z_n = \Omega \, \pder{}{\Omega} \ln Z \simeq \frac{\beta\hbar\Omega}{\sqrt{1+(d/\hbar\Omega)^2}}, \]
		where we have used the fact that $\tanh(\half\beta\Delta)\simeq1$ in the 
		limit $\beta\rightarrow\infty$. 
	}
	
	In the case of a symmetric system,
	$K_0(\tau_\text{k}) = K_\ell(\tau_\text{k}) = K_r(\tau_\text{k})$,
	$d=0$ and $\Delta = 2\hbar \Omega$, where $\Omega\simeq\lim_{\tau_\mathrm{k}\rightarrow\infty} K_1'(\tau_\mathrm{k})/K_0(\tau_\mathrm{k})$,
	and thus
	one recovers the 
	standard instanton result for the tunnelling splitting. \cite{Uses_of_Instantons,Benderskii,tunnel,InstReview}
	It therefore turns out that, although it was necessary to derive the more general asymmetric theory in a subtly different way,
	the only 
	significant change to the final expression for the tunnelling matrix element is 
	simply that the denominator should be replaced by the geometric mean of the two non-tunnelling kernels.

	\subsection{Ring-Polymer Instanton Theory}
	
	The instanton tunnelling pathways
	are uniquely defined as minimum-action paths connecting the two wells \cite{Uses_of_Instantons} and are equivalent to Newtonian trajectories travelling on the upside-down potential-energy surface. \cite{Miller1971density}
	However, in general multidimensional systems,
	these 
	pathways are not known analytically. 
	We thus employ the ring-polymer instanton approach \cite{tunnel,Perspective,InstReview} in order to obtain a practical method which can be used to 
	locate the pathways and evaluate 
	the level splitting
	in full-dimensional molecular systems.
	
	In order to compute
	$\theta(\tau_\text{k})$,
	we only need to consider the contribution from a single kink or paths with no kinks
	and thus do not require a cyclic ring polymer but only a segment of it of length $\tau_\text{k}$.
	The path-integral representation of the propagator is discretized into $N$ imaginary-time intervals
	using $N-1$ beads, $x_i$, which each correspond to a replica of the $f$-dimensional system.
	The imaginary-time action (also known as the Euclidean action) associated with this path is then \cite{Feynman}
	\begin{align} \label{S}
	S(\bm{x}) = \sum_{i=1}^N \frac{m}{2\tau_N}\left|x_i - x_{i-1}\right|^2
	+ \sum_{i=1}^N \frac{\tau_N}{2} \left[V(x_{i-1}) + V(x_{i})\right],
	\end{align}
	where $\tau_N = \tau_\text{k}/N$, $\bm{x} = \{x_1, \cdots, x_{N-1}\}$, and $x_0$ and $x_N$ are the fixed end points, equal to $x_\ell$ or $x_r$ as appropriate.
	
	Performing the path integral within the steepest-descent approximation
	gives the single-kink contribution: \cite{tunnel,InstReview}
	\begin{align} \label{K1}
	K'_1(\tau_\text{k}) = \frac{1}{\tau_N}\left(\frac{m}{2\pi\tau_N\hbar}\right)^{f/2} \sqrt{\frac{S(\tilde{\bm{x}})}{2\pi\hbar}} \, [\mathrm{det}'(\mathbf{J})]^{-1/2} \, \eu{-S(\tilde{\bm{x}})/\hbar} ,
	\end{align}
	where $\mathbf{J} = \frac{\tau_N}{m}\bm{\nabla}^2 S(\tilde{\bm{x}})$,
	$\bm{\nabla}$ is a derivative with respect to $\bm{x}$,
	and the primed determinant signifies that the zero-frequency mode associated with the permutational invariance of the instanton is not included in the product of eigenvalues.\cite{tunnel,InstReview}
	The instanton path, $\tilde{\bm{x}}$, is optimized so as to minimize the action, $S(\tilde{\bm{x}})$.
	Thus, the implementation of this method is similar to that of ring-polymer instanton rate theory,\cite{Miller1975semiclassical,RPInst,Andersson2009Hmethane,Althorpe2011ImF,Rommel2011locating,Perspective,InstReview} except for the main difference that the tunnelling pathway is not a saddle point but the minimum of the action. 
	
	Similarly, the contribution from the non-tunnelling paths starting and ending in the left well is
	\begin{align}
	K_\ell(\tau_\text{k}) = \left(\frac{m}{2\pi\tau_N\hbar}\right)^{f/2}[\det (\mathbf{J}_\ell)]^{-1/2},
	\end{align}
	where the fluctuation matrix is $\mathbf{J}_{\ell} = \frac{\tau_N}{m}\bm{\nabla}^2 S (\bm{x}_{\ell})$,
	and $\bm{x}_{\ell}$ is a collapsed path which has all its beads located at the bottom of the left well.
	The expressions for $K_r(\tau_\text{k})$ and $\mathbf{J}_r$ are equivalently defined in terms of a path collapsed in the right well.%
	\footnote{Given the well frequencies, $\mathbf{J}_\ell$ and $\mathbf{J}_{r}$
		can be evaluated analytically.  See Eq.~(16) of Ref.~\citenum{InstReview}.}
	
	Taking the ratio of kernels as defined in \eqn{eq:theta} thus gives 
	\begin{align}
	\label{eq:rp-inst}
	\theta(\tau_\text{k}) &=
	\lim_{N \rightarrow \infty}
	\frac{1}{\Phi}\sqrt{\frac{S(\tilde{\bm{x}})}{2\pi\hbar}} \, \eu{-S(\tilde{\bm{x}})/\hbar},
	\end{align}
	where
	\begin{align}
	\Phi = \tau_N \left[ \frac{\det'(\mathbf{J})}{\sqrt{
			\det(\mathbf{J}_\ell) \det(\mathbf{J}_{r})
	}}\right]^{1/2}
	\end{align}
	and we have indicated explicitly that these expressions must be converged with respect to the number of beads.
	
	
	For a calculation converged with respect to $\tau_\mathrm{k}$ and $N$, the action will be independent of the location of the centre of the kink.
	Therefore, as for a symmetric system, $\mathbf{J}$ still has a zero mode, which is integrated out in the standard way. \cite{Uses_of_Instantons,Benderskii,tunnel,InstReview}
	However, unlike for a symmetric system, $\Phi$ will depend weakly on the location of the kink as the fluctuations in the two wells are different.
	To be consistent with the derivation presented above,
	we will thus require that the kink is centred such that it reaches its maximum potential energy at imaginary time $\tau_\mathrm{k}/2$ (at least to the nearest bead).
	In our experience, this is not difficult to achieve.  For instance, after an optimization, one can easily correct a path by removing beads from one end and adding them to the other.%
	\footnote{Alternatively, one could redefine the 
		denominator in $\Phi$
		in an asymmetric way, 
		by changing $\sqrt{\det(\mathbf{J}_\ell)\det(\mathbf{J}_r)}$ to $\det(\mathbf{J}_\ell)^\gamma \det(\mathbf{J}_r)^{1-\gamma}$
		where $\gamma$ is the fraction of imaginary time (relative to $\tau_\mathrm{k}$)
		at which the kink reaches its maximum potential energy.  The fluctuations of the beads located in the two wells on either side of the kink will then cancel with the denominator in such a way to ensure that the result converges with respect to $\tau_\mathrm{k}$.}

	In order to apply this formulation to molecular systems in full-dimensionality,
	translational and rotational degrees of freedom are treated in the usual way \cite{InstReview}
	by removing their corresponding modes from the product of the eigenvalues when computing the determinants.
	%
	Finally, one does not typically know the relative orientation of $x_\ell$ and $x_r$. \cite{tunnel}
	In this case, the whole path including its end points, $\{x_0,\dots,x_N\}$, must be optimized, for instance using the l-BFGS algorithm \cite{Liu1989lBFGS}
	or recently developed specialized methods based on extensions of the the nudged-elastic band approach. \cite{Einarsdottir2012path,Cvitas2016instanton,Cvitas2018instanton,Erakovic2020instanton,pentamer,WaterChapter}
	One could also avoid numerical diagonalization of $\mathbf{J}$ by solving a related asymmetric Riccati equation.\cite{Erakovic2020instanton}

	\section{Results}
	
	We will apply the new theory to one- and two-dimensional double-well models to benchmark the approach against exact quantum-mechanical results.
	Finally, we apply the method to a set of isotopically substituted malonaldehyde molecules in full dimensionality and compare with experimental measurements.
	
	\subsection{1D Asymmetric Double Well}
	\label{subsec:1d-asym}
	
	
	In order to benchmark our new theory in the simplest asymmetric case,
	we modify the well-known quartic double-well potential, \cite{Polyakov1977instanton,ABCofInstantons,Benderskii}
	which has previously been used to test the standard ring-polymer instanton theory. \cite{tunnel}
	To create an asymmetric double-well potential, the symmetric version can be multiplied by a polynomial $P(x) = a x^2 / x_0^2 + b x/x_0 + c$.  The overall potential will remain a double well if the the parameters are chosen such that $P(x)$ has only imaginary roots.
	We also choose $c=1$ such that it does not significantly change the barrier height.
	We therefore define the one-dimensional potential as
	\begin{align}
	V(x) = V^\ddag \left(\frac{x^2}{x_0^2} - 1\right)^2 \left(a\frac{x^2}{x_0^2}+b\frac{x}{x_0}+1\right).
	\label{V1D}
	\end{align}
	This model describes a system 
	with two minima of equal potential but
	with one well sightly broader than the other%
	\footnote{Note that it is necessary for the potential function to be at least a sextic polynomial in order to demonstrate this behaviour.}
	and is depicted in Fig.~\ref{fig:potential}.
	In particular, the minima are located at $x=\pm x_0$ with $V(\pm x_0)=0$ and 
	the harmonic frequencies are $\omega_\ell = \omega_0 \sqrt{1+a-b}$ and $\omega_r = \omega_0 \sqrt{1+a+b}$,
	where $\omega_0 = \sqrt{8 V^\ddag/mx_0^2}$.
	Hence, $E_0$ and $d$ are easily found.
	
	Note that in the case that both $a\ll1$ and $b\ll1$, then $d \approx \frac{1}{4} b \hbar \omega_0$,
	such that we can identify $b$ as a parameter which controls the asymmetry.
	In order to ensure that the model has only two minima, we additionally require that $a > b^2/4$, although the parameter $a$ does not otherwise have a significant qualitative effect on the shape of the potential.
	In our tests, we therefore use $a=b$, although this choice has little bearing on our conclusions,
	and vary the value of $b$ to achieve various degrees of asymmetry.
	The remaining parameters are chosen 
	to match the symmetric ($b=d=0$) case tested 
	in Ref.~\onlinecite{tunnel}.
	In particular, reduced units are used in which $m=1$ and $\hbar=1$.
	The barrier height is varied via $V^\ddag$ but
	we choose $x_0 = 5\sqrt{V^\ddag}$ such that $\omega_0$ is fixed and hence $E_0\approx0.283$ in each case.

	
	\begin{table}
		\centering
		\caption{
			Results are presented for the one-dimensional double-well model [\eqn{V1D}] with three different barrier heights, $V^\ddag$, in three parameter regimes:
			the top section with $a=b$ chosen to demonstrate a weak asymmetry ($d\ll\hbar\Omega$);
			the middle section with medium asymmetry ($d\sim\hbar\Omega$);
			and the bottom section with large asymmetry ($d\gg\hbar\Omega$).
			In each case, $x_0=5\sqrt{V^\ddag}$ is fixed to keep $\omega_0$ constant.
			$\Omega$ is approximated by $\theta(\tau_\mathrm{k})$ [\eqn{eq:rp-inst}]
			and
			$\Delta_{\text{inst}} = 2\sqrt{d^2 + (\hbar \Omega)^2}$,
			whereas the quantum-mechanical benchmark,
			$\Delta_\text{QM}$, is calculated directly from DVR\@.
			Powers of 10 are given in parentheses.
		}
		\label{tab:1D}
		\small
		\begin{ruledtabular}
			\begin{tabular}{cd{1.4}d{1.7}d{1.6}d{1.6}d{1.5}r}
				$V^\ddag$ & \ctitle{$b$} &\ctitle{$d$} & \ctitle{$\hbar\Omega$} & \ctitle{$\Delta_{\mathrm{inst}}$} & \ctitle{$\Delta_{\mathrm{QM}}$} & \ctitle{Error}\\
				\hline
				0.5 & 1.(-3)  & 1.41(-4) & 1.52(-2) & 3.05(-2) & 2.25(-2) & 35\%\\
				1   & 1.(-5)  & 1.41(-6) & 1.93(-4) & 3.86(-4) & 3.42(-4) & 13\% \\ 
				2   & 1.(-9)  & 1.41(-10) & 2.20(-8) & 4.39(-8) & 4.15(-8) & 6\% \\
				\hline
				\hline
				0.5 & 1.(-1)  & 1.35(-2) & 1.76(-2) & 4.39(-2) & 3.13(-2) & 40\%\\
				1   & 1.(-3)  & 1.41(-4) & 1.93(-4) & 4.79(-4) & 4.27(-4) & 12\%  \\
				2   & 1.(-7)  & 1.41(-8) & 2.20(-8) & 5.22(-8) & 4.95(-8) & 6\%\\
				\hline \hline
				1 & 1.(-1) & 1.35(-2) & 2.11(-4) & 2.70(-2) & 2.39(-2) & 13\%\\ 
				2 & 1.(-5) & 1.41(-6) & 2.19(-8) & 2.83(-6) & 2.70(-6) &  5\%\\ 
			\end{tabular}
		\end{ruledtabular}
	\end{table}
	
	Table~\ref{tab:1D} presents the numerical results of the new asymmetric instanton theory
	and compares them with 
	benchmark quantum-mechanical 
	results.
	{The latter are} obtained from a numerical solution of the Schr\"odinger equation
	using the discrete variable representation (DVR),\cite{Meyer1970DVR,Light1985DVR}
	{in which the energy splitting is calculated from the difference between the two lowest eigenvalues.}
	The ring-polymer instanton calculations were found to be converged in each case with $\tau_\text{k} = 30$ and $N = 128$.
	We compare three regimes, one where $b$ is chosen to describe a weak asymmetry ($d \ll \hbar\Omega$), one with medium asymmetry ($d \sim \hbar\Omega$) and one with high asymmetry ($d \gg \hbar\Omega$).
	
	Ring-polymer instanton theory for the fully symmetric case (with $a=b=0$ and hence $d=0$) has already been tested in Ref.~\onlinecite{tunnel} and can even be evaluated analytically. \cite{ABCofInstantons}
	It was shown that the instanton approximation is more accurate for higher barriers, where $\Omega$ is small, and less accurate for lower barriers, where $\Omega$ approaches the size of $\omega_0$.
	This is expected of a semiclassical method which gives only the leading exponential behaviour \cite{Kleinert}
	and is due to the accuracy of the steepest-descent approximation, 
	which improves as the ratio between $\hbar$ and fluctuations of the action decreases.
	
	In the case of very weak asymmetry, 
	the predictions for the level splitting are almost identical to those of the symmetric case, \cite{tunnel}
	a pattern which is reproduced in the quantum results.
	This is because $\Delta$ is dominated by the tunnelling contribution $\hbar\Omega$, which is barely affected by the slight asymmetry added to the system.
	On the other hand, for high asymmetry, the level splitting is dominated by $d$ and again gives good agreement with the quantum mechanics.\footnote{We have not presented results for the $V^\ddag=0.5$ case of high asymmetry, for which $d$ would necessarily be similar in size to the vibrational spacing ($\hbar\omega_0$) and thus contrary to our initial assumption that the system is well described by a two-level effective Hamiltonian.}
	However, for a medium asymmetry such that $d$ and $\hbar\Omega$ are of comparable size,
	the tunnelling matrix element, $\Omega$, decreases with increasing barrier height as before, but now
	there is also a significant contribution from $d$ to $\Delta$.
	Importantly, in each case our asymmetric instanton calculations still give accurate predictions for the level splitting
	with errors which are similar to those found in the symmetric system. 
	Like them, the error decreases with increasing barrier height and due to the properties of the semiclassical approximation would become exact in the $\hbar\rightarrow0$ limit.
	
	
	
	
	{Finally, we calculated the mixing angle, $\phi$, from DVR (by defining the populations in terms of the integrals of the square of the first excited wavefunction to the left and right of its node)
		and compared them with the values predicted from instanton theory, which are evaluated directly from the results presented in Table~\ref{tab:1D}\@.
		Both methods are in agreement that
		in the weak asymmetry cases, the angle is greater than $89^\circ$,
		and in the high asymmetry cases, the angle is less than $1^\circ$.
		For the medium asymmetry case, the DVR values of $\phi$ were calculated as $51.4^\circ$, $53.3^\circ$ and $57.0^\circ$ for the three systems tested in order of increasing barrier height, whereas the predictions of instanton theory are $52.3^\circ$, $53.8^\circ$ and $57.2^\circ$.
		These are in excellent agreement and show only a tiny error which decreases as the barrier height is raised.
	}

	\subsection{2D Asymmetric Mode-Coupling Potential}
	
\begin{figure}
		\centering
		\includegraphics[width=\linewidth,keepaspectratio]{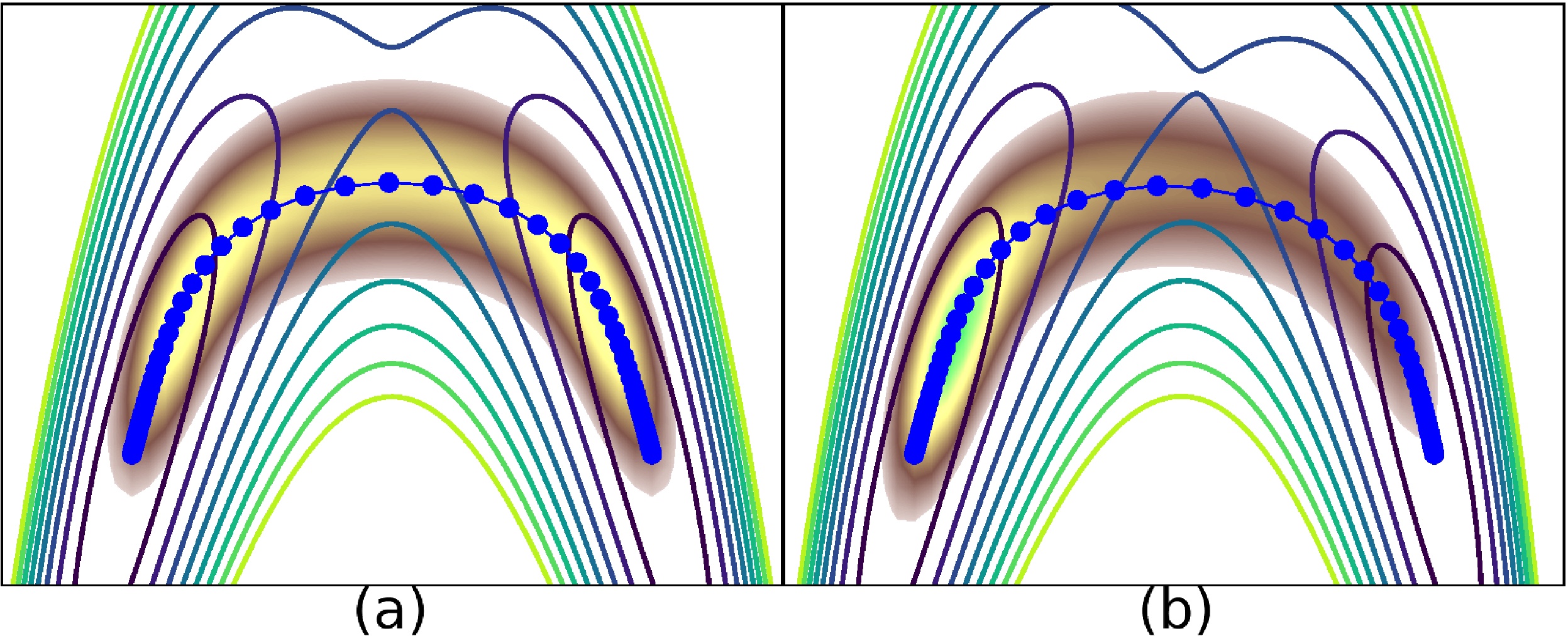} 
		\caption{Contour plots of the (a) symmetric and (b) exaggerated asymmetric mode-coupling potentials.
			In each case the ring-polymer instanton path is shown (by the blue curve connected by the ring-polymer beads represented by the blue circles).
			Also indicated is the norm of the quantum flux or current density (shown by yellow shading).
			\cite{*[{This is defined by $|F|$ where
					$F=\frac{\hbar}{m}\Imag[\psi^*\nabla\psi]$,
					$\psi = (\psi_- + \iu \psi_+)/\sqrt{2}$
					and the stationary wavefunctions $\psi_\pm$ are obtained from DVR\@.
					This illustration was inspired by a similar figure presented by }] [{}] KaesterFlux}
			The instanton path follows the regions of high flux density and thus clearly gives a good description of the dominant tunnelling process
			in both the symmetric and asymmetric case.
		}
		\label{fig:2d-pes}
	\end{figure}

	
	A commonly used simple model of two-dimensional tunnelling is provided by the
	symmetric mode-coupling potential. 
	\cite{Benderskii}
	%
	In this work, we generate an asymmetric mode-coupling potential by 
	following the approach used in Sec.~\ref{subsec:1d-asym},%
	\footnote{Similar two-dimensional models have been used by Bosch et al.\cite{Bosch1992malonaldehyde} to study isotopic substitutions in malonaldehyde, but are not appropriate for our needs as the zero-point energy difference is built into the model giving wells of slightly different energies.}
	\begin{align}
	V_{\mathrm{AMC}}(x,y) = \frac{1}{8}(x-1)^2(x+1)^2(ax^2 + bx + c) \nonumber \\ 
	+ \frac{\omega_y^2}{2}\left[y + \alpha\left(x^2-1\right)\right]^2 ,
	\label{AMC}
	\end{align}
	where $\omega_y$ is the frequency in the $y$ direction and $\alpha$ is a measure of coupling strength.
	%
	As before we choose $c=1$ and $a=b$ but vary the value of $b$ to generate systems in different regimes.
	The remaining parameters were chosen following \Ref{NakamuraBook} as $m = \alpha=1$, $\omega_y = 0.8$ and $\hbar = 0.04$.
	The AMC potential and its symmetrized version are shown in Fig.~\ref{fig:2d-pes}. It should be noted that in 
	order to better illustrate the asymmetry of the potential-energy surface to the reader,
	we have chosen such a large value for $a=b$ 
	that the level splitting would be completely dominated by the zero-point energy difference.
	Numerical calculations were however performed with smaller values
	of $b$ to demonstrate the most interesting regimes. 
	
	\begin{table} 
		\centering
		\caption{Level splittings obtained from the instanton method ($\Delta_{\mathrm{inst}}$) as well as from exact quantum mechanics via DVR ($\Delta_{\mathrm{QM}}$) for the 2D symmetric and asymmetric mode-coupling potential [\eqn{AMC}].
			The parameter $b$ was varied to control the amount of asymmetry, although the average well frequency remained approximately constant such that $E_0\approx 0.0482$ in each case.
			The ZPE difference, $d$, is obtained from the harmonic approximation of the two wells
			and tunnelling matrix element, $\hbar\Omega$, is estimated from instanton theory.
			All quantities are given in reduced units with powers of 10 indicated in parentheses.}
		\label{tab:asmc}
		\small
		\begin{ruledtabular}
			\begin{tabular}{cd{1.7}cccr}
				$b$ & \ctitle{$d$} & $\hbar\Omega$ & $\Delta_{\mathrm{inst}}$ & $\Delta_{\mathrm{QM}}$ &Error\\
				\hline 
				0           & \ctitle{$0$}  & 2.48(-9) & 4.96(-9) & 4.55(-9) & 9\% \\
				$0.\Bar{3}(-8)$ & 2.49(-11) & 2.48(-9) & 4.96(-9) & 4.55(-9) & 9\% \\
				$0.\Bar{3}(-6)$ & 2.49(-9)  & 2.48(-9) & 7.03(-9) & 6.59(-9) & 6\% \\
				$0.\Bar{3}(-4)$ & 2.49(-7)  & 2.47(-9) & 4.98(-7) & 4.76(-7) & 5\% \\
			\end{tabular}
		\end{ruledtabular}
	\end{table}
	
	We present the level splittings calculated for varying degrees of asymmetry in Table \ref{tab:asmc}.
	The instanton results were converged using $N=1024$ and $\tau_k=100$. 
	The trends for the 2D case mirror those of the 1D case.
	In particular, the prediction for the case of weak asymmetry is almost identical to the fully symmetric case.
	For medium asymmetry, both $d$ and $\Omega$ make significant contributions to $\Delta$.
	Finally for the most asymmetric case where $d \gg \hbar\Omega$, 
	the level splitting is completely dominated by the zero-point energy and $\Delta\approx 2d$ as expected.
	It can be seen that the agreement between the instanton approach and the quantum-mechanical benchmark is excellent, with errors under 10\% in all cases.

	\subsection{Isotopic Substitutions in Malonaldehyde} \label{subsec:malonaldehyde}
	The proton transfer in malonaldehyde has been used in many studies as a benchmark system for calculating tunnelling splittings using various theoretical approaches,
	such as
	reduced-dimensionality quantum mechanics \cite{Bosch1990malonaldehyde,Bosch1992malonaldehyde,Feng20172DMalon}
	the multiconfigurational time-dependent Hartree (MCTDH) method,\cite{Schroeder2011malonaldehyde, Coutinho2004malonaldehyde, SchroederMCTDH2014, Hammer2011malonaldehyde, manthe2012} 
	and diffusion Monte Carlo (DMC),\cite{Wang2008malonaldehydePES, Tew2014PES}
	as well as various flavours of instanton theory \cite{Milnikov2003,*Milnikov2004, tunnel, MUSTreview, Cvitas2016instanton,Cvitas2018instanton,Erakovic2020instanton}
	and other reaction-path approximations. \cite{Bicerano1983malonaldehyde,*Carrington1986malonaldehyde,*Ruf1988malonaldehyde,Yagi2001malonaldehyde,Tautermann2002malonaldehyde,Wang2008malonaldehydeSplit,*Wang2013Qim,Burd2020tunnel}
	The majority of these studies, however, do not examine the effects of isotopic substitution, or at best,
	only deuterate one hydrogen atom (called \ce{D6} in the notation employed below).

	Experimental studies by Baughcum et al.\cite{Baughcum1981malonaldehyde, Baughcum1984malonaldehyde} examine isotopic substitution in more detail,
	and in this work we shall compare our results to their measurements.
	We thus evaluate the level splitting for a number of isotopically symmetric and asymmetric species of malonaldehyde in full dimensionality.
	Following \Ref{Baughcum1981malonaldehyde}, we shall label the atoms as indicated in Fig.~\ref{fig:d7}
	and denote the various isotopomers by stating only the substituted atoms labelled with a subscript.
	
	\subsubsection{Instanton calculations}
	
	The molecular system is described by the ground-state
	Born--Oppenheimer potential-energy surface (PES)\@.
	We have employed two PESs for these calculations; one constructed by Mizukami et al. \cite{Tew2014PES}  (referred to as PES1) and one by Wang et al. \cite{Wang2008malonaldehydePES} (referred to as PES2)\@.
	Both are fitted to
	high-level electronic-structure calculations based on coupled-cluster theory. 
	The more recent PES1 also employs explicit correlation using F12* theory\cite{f12-david} and reports a lower fitting error
	and is thus expected to be the more accurate of the two, although the differences in this case are not particularly dramatic. 
	{For instance, the barrier heights for PES1 and PES2 are 4.03 and 4.11~$\mathrm{kcal\,mol^{-1}}$ respectively.} 
	
	
	Instanton optimizations were carried out by minimizing the action in Cartesian coordinates as outlined in Refs.~\citenum{tunnel} and \citenum{InstReview}.
	In addition,
	we must analyse the minimum-energy configurations on both sides of the barrier to obtain their harmonic frequencies.
	With this information we can compute the necessary quantities required 
	to evaluate $\theta(\tau_\mathrm{k})$ and hence predict $\Omega$.
	Calculations were performed with increasing values of $N$ and $\tau_\mathrm{k}$ until $\theta(\tau_\mathrm{k})$ was converged.
	This was achieved at $N=1024$ and an effective ``temperature'' of $\hbar/(k_\mathrm{B}\tau_\mathrm{k}) = 30$\,K (i.e.\ $\tau_\mathrm{k} \approx 10\,000$\,a.u.).
	Finally, $d$ and $\Omega$ were combined to predict $\Delta$.
	
	
	The symmetric parent molecule has been used widely as a benchmark to test the accuracy of quantum dynamics methods and potential-energy surfaces
	and thus results obtained with various methods are available.
	Two different MCTDH calculations using PES2 obtain values for $\Delta$ of 23.4 and 23.8 cm$^{-1}$ for the parent configuration. \cite{Hammer2011malonaldehyde,Schroeder2011malonaldehyde}
	{These are in relatively good agreement with various
		calculations (also using PES2) based on 
		fixed-node DMC
		or the difference between the DMC ZPE and a local ZPE analysis,}
	which obtained a tunnelling splitting in the range 21--22 $\mathrm{cm}^{-1}$ for the parent molecule and 2--3 $\mathrm{cm}^{-1}$ for the symmetric (\ce{D6}) isotopomer (both with statistical uncertainties of between 2 and $3\,\mathrm{cm}^{-1}$). \cite{Wang2008malonaldehydePES}
	Using the same PES, our ring-polymer instanton results give values of 24.9 and 3.29 cm$^{-1}$ for these cases.
	Note that, a similar calculation based on a different implementation of instanton theory \cite{Cvitas2016instanton} 
	reports tunnelling splittings within $0.1\,\mathrm{cm}^{-1}$ of those calculated in this work.
	Results generated using the newer PES1 will not necessarily be the same
	due to slight differences in the level of electronic-structure employed as well as the fitting method when constructing the PESs.
	Using PES1, the DMC results of Mizukami et al. \cite{Tew2014PES} are $21.0\pm0.4\,\mathrm{cm}^{-1}$ for the parent molecule
	and $3.2\pm0.4\,\mathrm{cm}^{-1}$ for the (\ce{D6}) isotopomer.
	With our ring-polymer instanton method, 
	we obtain values of 19.3 and 2.69 cm$^{-1}$.
	Therefore in both cases, our instanton predictions are in good agreement with the (far more computationally intensive) MCTDH and DMC methods,
	with an error just under 10\% for the parent molecule, which 
	is remarkably no less accurate than for the simple low-dimensional models.
	Our predictions using the two different PESs are noticeably different, with one tending to overpredict and the other underpredict the experimental splittings.
	{This discrepancy in the results from different PESs is an indication of how sensitive the calculations are to subtle details of the PES\@.
		Due to these uncontrollable factors as well as the inherent approximations of instanton theory, it will be impossible to quantitatively reproduce the experimental tunnelling splittings to high accuracy.
		However, one may still be able to explain and predict the trends obtained upon isotopic substitutions, as we will now show.}

	\begin{figure}
		\centering
		\includegraphics[width=0.7\linewidth]{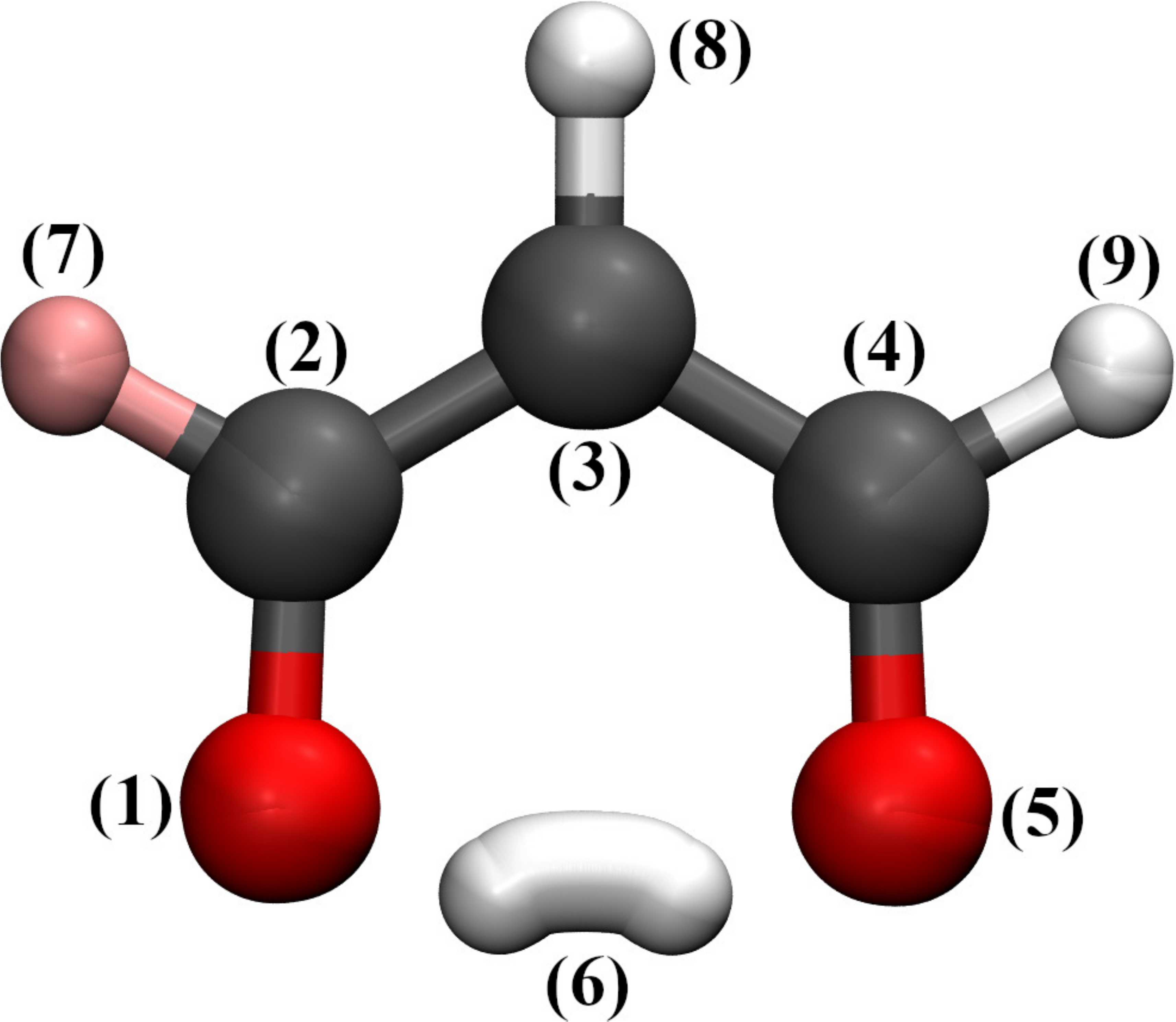} 
		\caption{Configuration of the ring-polymer instanton beads for the asymmetric (\ce{D7}) isotopomer of malonaldehyde, \ce{C3H4O2}, in which the substituted deuterium is coloured pink.
			Other isotopomers show qualitatively similar behaviour and in each case the
			\ce{H6}/\ce{D6} atom dominates the tunnelling dynamics and the
			molecule remains planar along the whole pathway.
			This figure also defines our labelling scheme, which follows that used by Baughcum et al.\cite{Baughcum1981malonaldehyde}
		}
		\label{fig:d7}
	\end{figure}{}

	The calculated level splittings for various isotopic substitutions of malonaldehyde are presented in Table~\ref{tab:all-malonaldehyde} along with experimental measurements where available. \cite{Baughcum1984malonaldehyde}
	The instanton pathway shows a similar behaviour in each case and is not significantly affected by whether the isotopomer is symmetric or asymmetric.
	A representative instanton pathway is presented in Fig.~\ref{fig:d7} which
	provides an intuitive picture of the tunnelling mechanism in malonaldehyde.
	For instance, it can be observed that the \ce{H6} atom 
	is the only one directly involved in the proton transfer process and 
	makes the largest contribution to the tunnelling path. 
	We can quantify this in the same manner as we have done in our previous work\cite{Muonium} by evaluating the squared mass-weighted path length, which is in this case proportional to the action, $S$.\footnote{This is because the action for a path with zero total energy is equal can be rewritten in the time-independent form \mbox{$S=\oint p\,\rmd x = \sum_{i=1}^N \frac{m}{\tau_N} |x_i - x_{i-1}|^2$} using mass-weighted coordinates.}
	From this, we find that atom 6 contributes about 70\% of the total action, whether it is deuterated or not, although the overall path length (and hence the action) does of course increase significantly when a \ce{D6} substitution is introduced.
	Tunnelling is thus strongly dependent on the isotope of this particular atom, 
	as our results shall demonstrate.

	This atom continues to dominate the action for the instantons associated with every other isotopic substitution under study.
	While this might seem to imply that a reduced-dimensionality approach based on the dynamics of this single atom would yield decent results,
	previous work \cite{Yagi2004malonaldehyde,formic} has shown that in general it is necessary to include all degrees of freedom (even those not involved in the tunnelling pathway)
	to obtain an accurate value for $\Omega$ and thus the level splitting $\Delta$.  
	
	\subsubsection{Discussion of the Results}
	
	We first discuss the results for the symmetric isotopomers, which are presented in the top half of Table~\ref{tab:all-malonaldehyde}.
	In these cases $d=0$ such that the level splitting simplifies to $\Delta=2\hbar\Omega$ and
	can thus be called a tunnelling splitting.
	From the results, one can note that for as long as \ce{H6} is not substituted with deuterium, $\Delta$ remains similar to its value in the parent molecule.
	However, upon making the isotopic substitution \ce{D6}, $\Delta$ decreases dramatically, as expected.
	The experimental results also reveal that 
	introducing \ce{^13C} or \ce{^18O} isotopes
	causes a reduction in the tunnelling splitting.
	This trend is correctly captured by the instanton method due to the fact that
	the action increases slightly with the substitution of heavier isotopes into the molecular skeleton, 
	which thus results in a slight decrease in the predicted tunnelling splitting. 
	The presence of \ce{^18O} atoms appears to have a greater effect than \ce{^13C}, mostly because they are more strongly coupled to the transferred \ce{H6}/\ce{D6} atom and contribute about 20\% of the path action, whereas the C atoms contribute about 10\%.%
	\footnote{The fact that \ce{^18O} is two atomic mass units heavier than \ce{^16O} obviously also plays a role in exaggerating this effect, although we can show that it is not the dominant factor by calculating $\Omega$ in the presence of \ce{^17O} or \ce{^14C} isotopes, for which we still find that the O substitutions have a greater effect.}
	However, isotopic substitution of the other three H atoms (i.e.\ \ce{D7}, \ce{D8} and \ce{D9}) appears to be more complex 
	because they are only very weakly coupled to the tunnelling pathway and contribute less than 1\% to the path action.
	Substituting atoms for heavier isotopes is guaranteed to increase the action. 
	However, although they cause a tiny increase in the action, there is also a larger effect in the contribution to $\Phi$ from the fluctuations around the tunnelling path.
	Making the substitution \ce{D8} appears to consistently decrease the splitting slightly.
	On the other hand, according to some of the calculations, adding \ce{D7D9} may increase the splitting.
	This behaviour is also observed in the reported experimental splittings for isotopomers containing \ce{D6D7D9}. 
	Nonetheless, this sort of competing effect is hard to predict reliably with any {approximate} quantum-dynamics approach, and we see that the two different PESs give opposite trends for \ce{D7D9} compared with the parent molecule.
	\begin{table}
		\caption{Level splittings, $\Delta$,
			and its contributions from ZPE ($d$) and tunnelling ($\hbar\Omega$)
			for various symmetric and asymmetric isotopic substitutions of malonaldehyde
			calculated from instanton theory using two different PESs.
			The level splittings of the symmetric and asymmetric isotopomers are separated by the double-lines, 
			with the upper half being the symmetric isotopomers and the bottom half being the asymmetric isotopomers.
			All quantities are reported in $\mathrm{cm}^{-1}$.}
		\label{tab:all-malonaldehyde}
		\begin{ruledtabular}
			\begin{tabular}{l d{1.2}d{1.2}d{2.2} c d{1.2}d{2.2}d{2.1} c d{1.2}} 
				\multirow{2}{*}{Isotopomer} & \multicolumn{3}{c}{PES1 \cite{Tew2014PES}} && \multicolumn{3}{c}{PES2\cite{Wang2008malonaldehydePES}} && \multicolumn{1}{c}{Expt.\cite{Baughcum1984malonaldehyde}} \\ \cline{2-4} \cline{6-8} \cline{10-10} 
				&   \ctitle{$|d|$} & \ctitle{$\hbar\Omega$} & \ctitle{$\Delta$} &&  \ctitle{$|d|$} & \ctitle{$\hbar\Omega$} & \ctitle{$\Delta$} && \ctitle{$\Delta$}\\
				\hline
				Parent      & 0 & 9.65 & 19.3 && 0 & 12.5 & 24.9 && 21.6 \\ 
				\ce{(D7D9)}   & 0 & 9.90 & 19.8 && 0 & 12.4 & 24.8 && -\\
				\ce{(D7D8D9)} & 0 & 9.75 & 19.5 && 0 & 11.9 & 23.7 && -\\
				\ce{(D8)}     & 0 & 9.50 & 19.0 && 0 & 11.8 & 23.6 && -\\
				\ce{(D6)}     & 0 & 1.35 & 2.69 && 0 & 1.65 & 3.29 && 2.92 \\
				\ce{(D6D7D9)} & 0 & 1.41 & 2.82 && 0 & 1.65 & 3.29 && 2.94 \\
				\ce{(D6D7D8D9)}   & 0 & 1.36 & 2.71 && 0 & 1.55 & 3.10 && 2.91 \\
				\ce{(D6D8)}   & 0 & 1.32 & 2.64 && 0 & 1.55 & 3.09 && 2.88 \\
				\ce{(D6D8$\mathrm{^{13}C_{2}{}^{13}C_4}$)} & 0 & 1.30 & 2.60 && 0 & 1.52 & 3.05 && 2.86\\
				\!\footnotesize \ce{(D6D8$\mathrm{^{13}C_{2}{}^{13}C_3{}^{13}C_4}$)} & 0 & 1.30 & 2.59 && 0 & 1.52 & 3.03 && 2.83\\
				\ce{(D6D8$\mathrm{{}^{18}O_{1} {}^{18}O_5}$)} & 0 & 1.24 & 2.47 && 0 & 1.46 & 2.91 && 2.72\\
				\hline \hline
				\ce{(^13C2/4)} & 0.18 & 9.58 & 19.2 && 0.01 & 12.3 & 24.6 && -\\ 
				\ce{(D7/9)}     & 16.2 & 9.64 & 37.8 && 12.7 & 12.4 & 35.5 && -\\
				\ce{(D7/9D8)}   & 16.3 & 9.47 & 37.9 && 12.8 & 11.8 & 34.9 && -\\
				\ce{(D6D7/9)}   & 16.4 & 1.36 & 32.9 && 13.0 & 1.64 & 26.2 && -\\
				\ce{(D6D7/9D8)} & 16.5 & 1.33 & 33.2 && 13.1 & 1.55 & 26.3 && -\\
				\ce{(D6D8$\mathrm{{}^{13}C_{2/4}}$)} & 0.25 & 1.31 & 2.67 && 0.06 & 1.54 & 3.08 && 2.86 \\
				\ce{(D6D8$\mathrm{^{13}C_{2/4}{}^{13}C_3}$)} & 0.28  & 1.30 & 2.67 && 0.11 & 1.53 & 3.07 && 2.84 \\  
				\ce{(D6D8$\mathrm{{}^{18}O_{1/5}}$)} & 2.29 & 1.28 & 5.25 && 2.19 & 1.50 & 5.31 && - \\ 

			\end{tabular}
		\end{ruledtabular}
	\end{table}

	Next, we discuss the level splittings for various asymmetric isotopic substitutions of malonaldehyde, which are presented in the bottom half of Table~\ref{tab:all-malonaldehyde}\@.
	These calculations were carried out using the generalized version of ring-polymer instanton theory developed in this paper.
	Here, three different regimes are demonstrated:
	$|d| \ll \hbar\Omega$ for isotopomers (\ce{^13C2/4}),  \ce{(D6D8$\mathrm{{}^{13}C_{2/4}}$)} and \ce{(D6D8$\mathrm{^{13}C_{2/4}{}^{13}C_3}$)};
	$|d| \sim \hbar\Omega$ for isotopomers (\ce{D7/9}), (\ce{D7/9D8}) and \ce{(D6D8$\mathrm{{}^{18}O_{1/5}}$)};
	and $|d| \gg \hbar\Omega$ for isotopomers (\ce{D6D7/9}) and (\ce{D6D7/9D8}).
	
	For the weakly asymmetric regime ($|d| \ll \hbar\Omega$), it can be observed that the level splitting is not significantly affected by the introduction of asymmetry, as was expected.
	This results in a level splitting for (\ce{^13C2/4}) 
	which is only slightly smaller than in the parent molecule.
	This behaviour is confirmed by the available experimental observations and can be attributed to the small skeletal rearrangement of the carbon atoms, which increase the action slightly on isotopic substitution, just as in the symmetric case.
	For the case of \ce{(D6D8$\mathrm{{}^{13}C_{2/4}}$)} and \ce{(D6D8$\mathrm{^{13}C_{2/4}{}^{13}C_3}$)}, the situation is not quite so simple as there are competing effects from an increased $|d|$ but a decreased $\Omega$,
	making it difficult to reliably predict the size of the level splittings relative to (\ce{D6D8}).
	It is however clear that \ce{(D6D8$\mathrm{{}^{13}C_{2/4}}$)} and \ce{(D6D8$\mathrm{^{13}C_{2/4}{}^{13}C_3}$)} have slightly larger splittings than \ce{(D6D8$\mathrm{^{13}C_{2}{}^{13}C_3{}^{13}C_4}$)} and \ce{(D6D8$\mathrm{^{13}C_{2}{}^{13}C_3{}^{13}C_4}$)} respectively
	due to their lower mass. 
	
	In the opposite extreme, $|d| \gg \hbar\Omega$, the level splitting is dominated by the asymmetry.
	The experiments \cite{Baughcum1981malonaldehyde} found no evidence of mixing between the two wells for (\ce{D6D7/9}) or (\ce{D6D7/9D8}).
	Our calculations confirm this scenario as the predicted mixing angles are about $5^\circ$ (PES1) or $7^\circ$ (PES2) in each case, which corresponds to a population ratio of more than 99:1. 
	{Note that in this case,
		the role of instanton theory is simply to confirm that the tunnelling effect is negligible.  The problem thus simplifies to one of finding the ZPE difference, for which alternative methods, which are more accurate than the harmonic approximation used here, are well established.\cite{Bowman2003multimode}}
	
	For the intermediate cases ($|d| \sim \hbar\Omega$), both asymmetry and tunnelling play a significant role in determining the level splitting.
	Just like with the symmetric isotopomers, 
	here it can be noted that when atom 6 is deuterated (i.e.\ isotopomers including \ce{D6}),
	a significant decrease in tunnelling (quantified by $\hbar\Omega$) can be observed. 
	Unfortunately no direct experimental values for the level splitting are reported for this set of isotopomers, although these are probably the most interesting cases.
	However, Baughcum et al.\cite{Baughcum1981malonaldehyde} do measure dipole moments and thereby roughly estimate the amplitude of the ground-state wavefunction in the two wells 
	for (\ce{D7/9}).
	Their estimate is equivalent to a mixing angle of $\phi\approx41^{\circ}$.
	Our instanton calculations predict a value of $31^\circ$ on PES1 or $44^\circ$ on PES2.
	This result is obviously quite sensitive to the PES, and the experimental estimate itself is only a rough approximation.
	However, it is nonetheless in agreement with the interpretation that the state is partially mixed,
	and demonstrates that our method is valid for this most difficult intermediate regime.
	

	
	Bosch et al. \cite{Bosch1990malonaldehyde,Bosch1992malonaldehyde}
	attempted to explain the experimental results using quantum-mechanical calculations based on a two-dimensional model system.
	Although the agreement with experiment is by no means quantitative [e.g.\ the predicted tunnelling splittings are 8.2 cm$^{-1}$ for the parent molecule and 0.3 cm$^{-1}$ for (\ce{D6})],\cite{Bosch1992malonaldehyde}
	they were able to roughly predict some of the trends, most notably the strong decrease in tunnelling splitting upon substituting with the \ce{D6} atom.
	Recently similar calculations \cite{Feng20172DMalon} using an improved two-dimensional surface based on PES2 have obtained more accurate results,
	but are still based on the reduced-dimensionality approximation, 
	which is known not to be reliable in general.\cite{formic}

	Bosch et al.\cite{Bosch1992malonaldehyde} also study the asymmetric \ce{(D6D7/9D8)} isotopomer and find that 
	the stationary states remain practically unmixed ($\phi\approx1^\circ$) due to strong asymmetry,
	which is in agreement with both our work and the experiment.
	However, for the (\ce{D7/9D8}) isotopomer, they predict a mixing angle of only 16$^\circ$, whereas our calculation gives a value at least twice as large (30$^\circ$ for PES1 and 43$^\circ$ for PES2).  This is quite a significant difference as it suggests that the ground-state population of the higher-energy well could be up to 6 times larger according to our theory.
	The main cause of the discrepancy is probably due to the fact that our PESs are based on much more accurate electronic structure than was available to Bosch et al.
	However, another important factor is the error introduced by their reduced-dimensionality approximation, which we are able to avoid by using a full-dimensional approach.
	Consequently, we conclude that the stationary states are much more mixed than was previously expected
	and that, even in these asymmetric systems, tunnelling still has a significant effect on the level splitting (such that it is about 5--9 cm$^{-1}$ larger than in the (\ce{D6D7/9D8}) isotopomer even though the zero-point energy differences are almost the same).
	In a similar way, the (\ce{D6D8{}^{18}O1/5}) isotopomer will be partially mixed due to its particularly large $d$ value caused by the isotopic substitution of the strongly coupled oxygen atom.
	Here the calculated mixing angle is $29^\circ$ on PES1 and $34^\circ$ on PES2 and it is therefore expected to demonstrate a particularly large level splitting relative to symmetric isotopomers with a \ce{D6} substitution.
	These predictions could potentially be confirmed by future experiments.
	
	
	Finally, we discuss the physical interpretation of our findings.
	As is clear from the theory presented in Sec.~\ref{sec:theory}, the level splitting is determined by two quantities $d$ and $\Omega$, which 
	quantify the effects of asymmetry and of tunnelling respectively.
	Our theoretical approach is able to calculate these two quantities independently of each other,
	which allows us to identify the relative magnitude of the two components.
	In contrast, experimental measurements of spectroscopic transitions can only provide direct information on the value of $\Delta$ and not the individual contributions.
	Based on their observations, Baughcum et al. \cite{Baughcum1981malonaldehyde} argue that tunnelling is ``quenched'' in asymmetrically substituted isotopomers.
	Whether or not this statement is consistent with our findings is perhaps somewhat a matter of semantics.
	In particular, if one quantifies tunnelling using the magnitude of $\Omega$, then our calculations demonstrate that $\Omega$ is approximately unaffected by asymmetric isotopic substitutions and has a significant dependence only on the \ce{D6} substitution.
	We do however find that the mixing angle may vary strongly 
	on the type of asymmetry
	in such a way that
	one can say that it is not \emph{tunnelling} but \emph{mixing} which is quenched.
	This subtly different interpretation remains consistent with all the experimental observations.

	We find that the asymmetric isotopomers of malonaldehyde exist in three regimes 
	which exhibit different quantum-dynamical behaviour.
	To some extent, one can design the isotopic substitutions to  
	demonstrate a ``sweet spot'' for which $|d|\sim\hbar\Omega$, and it would be interesting to test our predictions for the level splittings in this regime with new experiments. 
	We hope that our interpretation of the dynamics will be useful in explaining experimental observations
	and propose that the new method we have presented is used to help analyse and predict the level splittings in new molecules.

	\section{Conclusions}
	
	In this work we have generalized instanton theory to describe tunnelling in a certain type of asymmetric system.
	Although the final result for the level splitting is expressed by a trivial formula defined in terms of the zero-point energy difference, $d$, and the tunnelling matrix element, $\Omega$, 
	an instanton approach to compute the latter did not previously exist.
	Note that even if the system is almost symmetric such that $\Omega$ dominates the splitting,
	one could not have simply employed the standard instanton theory of Ref.~\onlinecite{tunnel}
	as this requires perfect symmetry in order to converge.
	
	Nonetheless, the final result shows that the new approach can be implemented in a very similar way to the standard ring-polymer instanton method and can thus be applied to complex molecular systems in full dimensionality in an efficient manner.
	The standard approach is currently available in i-PI \cite{iPI2} and MOLPRO \cite{Molpro} and can thus be used to study tunnelling in asymmetric systems too with very minor modifications to the code. 
	%
	%
	One could also employ machine-learning techniques as has been done with other instanton approaches \cite{GPR,Muonium,McConnell2019instanton} to allow efficient application with \textit{ab initio} electronic-structure calculations.
	
	We have applied the new methodology to benchmark one- and two-dimensional systems and shown that the method can predict level splittings in asymmetric systems just as accurately as standard instanton theory applied to symmetric systems.
	Finally, we have calculated the level splittings in various asymmetric isotopomers of malonaldehyde.
	For this molecule, we find that the value of $\Omega$ is only weakly dependent on the isotopes of all atoms except the transferred proton (\ce{H_6}/\ce{D_6}) and to a first approximation can be approximated by the value of the most similar symmetric isotopomer.
	The zero-point energy difference, $d$, is the only other
	factor which determines the level splitting
	and is controlled by the presence of isotopic substitutions in asymmetric positions.
	According to our definitions, the tunnelling effect and mixing angle can be studied independently.
	We find that although tunnelling (defined by $\Omega$) is not strongly affected by asymmetric substitutions,
	the mixing angle can be.
	The most interesting regime ($0\ll\phi\ll90^\circ$) occurs when $|d|\sim\hbar\Omega$,
	which is exhibited by asymmetric isotopomers with a transferring \ce{H6} atom and a \ce{D7/9} substitution or by those with a \ce{D6} atom and a \ce{^{18}O1/5} substitution.
	These simple rules would help assign and interpret any new experiments performed on similar molecules,
	and new calculations could easily be performed for other systems of interest.
	
	

	One could envisage further extensions of this methodology, for instance {by using graph-theory \cite{water} to} describe multi-well problems such as those occurring in water clusters. \cite{water,pentamer,hexamerprism,octamer,i-wat2,Erakovic2020instanton,WaterChapter}
	{It would clearly be possible to improve the overall prediction of the energy splitting (for some extra computational effort) by 
		combining the semiclassical instanton calculation of the tunnelling matrix element with} anharmonic calculations of the zero-point energy difference.{\cite{Bowman2003multimode}}
	{Furthermore,} to go beyond the semiclassical approximation {altogether}, {which is sometimes necessary for systems with low barriers,} it may be possible to apply
	path-integral sampling approaches
	\cite{Kuki1987proteins,*Alexandrou1988tunnelling,*Marchi1991tunnelling,Matyus2016tunnel1,*Matyus2016tunnel2,*Vaillant2018dimer}
	{to include anharmonic fluctuations around the instanton pathways}.
	
	{Finally, in future work we hope to build on this approach to describe tunnelling in more general systems, in which the wells have different potentials.}
	We note that 
	alongside its early development for the study of tunnelling in molecular systems, \cite{Miller1975rate,*Chapman1975rates}
	instanton theory was more widely employed in quantum field theory. \cite{Polyakov1977instanton,Uses_of_Instantons,ABCofInstantons}
	There may thus also be uses in statistical and particle physics
	for this asymmetric instanton theory. 
	

	

	
	\section*{Acknowledgements}
	The authors would like to thank Joel Bowman and David Tew for making their malonaldehyde PESs available.
	The authors acknowledge financial support from the Swiss National Science Foundation through Project 175696.

	\appendix
	\section{Useful results for the harmonic oscillator}
	\label{app:HO}
	Here we present formulas relating to path-integral and semiclassical results for the harmonic oscillator.
	In particular, we consider 
	the limit of long imaginary times, 
	$\omega_{\ell/r} \tau \gg 1$,
	such that the classical action for the harmonic oscillator can be approximated by \cite{Kleinert}
	\begin{align}
	S_0(x_a,x_b,\tau) \simeq \thalf m\omega_{\ell/r} \left[(x_b - x_{\ell/r})^2 + (x_a - x_{\ell/r})^2\right]
	\end{align}
	and the semiclassical kernel by
	\begin{align}
	K_0(x_a,x_b,\tau) \simeq
	\sqrt\frac{m\omega_{\ell/r}}{\pi\hbar\exp(\omega_{\ell/r}\tau)} \,
	\eu{-S_0(x_a,x_b,\tau)/\hbar} .
	\end{align}
	The special case of a kernel with both end-points at the bottom of the well is given by $S_0(x_{\ell/r},x_{\ell/r},\tau)=0$.
	
	When joining two kernels together, one finds
	\begin{align}
	K_0(x_a,x_c,\tau_1+\tau_2) &=
	\int K_0(x_a,x_b,\tau_1) K_0(x_b,x_c,\tau_2) \, \rmd x_b
	\nonumber \\ &= K_0(x_a,x_{\ell/r},\tau_1) \Xi_{\ell/r} K_0(x_{\ell/r},x_c,\tau_2) ,
	\end{align}
	where the term arising from the Gaussian integral is
	\begin{align}
	\label{eq:xi}
	\Xi_{\ell/r} &= \sqrt{2\pi\hbar}
	\left( \frac{\partial^2}{\partial x_b^2} [S_0(x_a,x_b,\tau_1) + S_0(x_b,x_c,\tau_2)] \right)^{-1/2}
	\nonumber
	\\& = {\sqrt\frac{\pi\hbar}{m\omega_{\ell/r}}} .
	\end{align}
	The same formulas can also be applied to join single-kink kernels together within the semiclassical approximation,
	and the multidimensional generalization can be found by simply taking products of these expressions over normal modes.

	\section*{Data Availability}
	The data that supports the findings of this study are available within the article.
	
	\section*{References}
	\bibliography{references,extra}
	
\end{document}